\documentclass[prl,twocolumn]{revtex4}
\usepackage{graphicx}
\usepackage{dcolumn}
\usepackage{bm}
\usepackage{amsmath}
\usepackage{txfonts}
\usepackage[scaled]{helvet}
\usepackage{color}
\usepackage[breaklinks=true,colorlinks,citecolor=blue,linkcolor=blue,urlcolor=blue]{hyperref}

\begin{document}

\title{Topological Valley Transport of Gapped Dirac Magnons in Bilayer Ferromagnetic Insulators}
\author{Xuechao Zhai$^{1,2}$ and Yaroslav M. Blanter$^2$}
\affiliation{$^1$New Energy Technology Engineering Laboratory of
Jiangsu Province $\&$ School of Science, Nanjing University of Posts
and Telecommunications (NJUPT), Nanjing 210023, China
\\$^2$ Kavli Institute of NanoScience, Delft University of Technology,
2628 CJ Delft, The Netherlands}

\begin{abstract}
Bilayer Heisenberg ferromagnetic insulators hold degenerate
terahertz Dirac magnon modes associated with two opposite valleys of
the hexagonal Brillouin zone. We show that this energy degeneracy
can be removed by breaking of the inversion symmetry ($\cal I$),
leading to a topological magnon valley current. We show furthermore
that this current leads to valley Seebeck effect for magnons and is
thereby detectable. We perform calculations in the specific example
of bilayer CrBr$_3$, where $\cal I$ can be broken by electrostatic
doping.
\end{abstract}

\maketitle

Recent discoveries of atomic-thick ferromagnetic (FM) insulators
\cite{Gong,Huang} represent a landmark for 2D fundamental and
applied physics. Stable long-range 2D FM order strongly relies on
the presence of magnetic anisotropy \cite{GongZhang,BurMand,GibKop}.
In these materials magnons, elementary excitations of magnetic
structure, usually have spectra with Dirac points. In particular,
magnons in two of the most popular 2D FM insulators have been
studied in more detail: gapless Dirac properties protected by
inversion symmetry $({\cal I})$ and
time-reversal-and-rotational symmetry $({\cal T}\mathcal{C}_r$) in
monolayer CrBr$_3$ \cite{PerBan} with no observable
Dzyaloshinskii-Moriya interaction (DMI), and topological gaps
induced by ${\cal T}\mathcal{C}_r$ breaking due to the
observable DMI in monolayer CrI$_3$ \cite{ChenChu}.
The operators ${\cal T}$, $\mathcal{C}_r$ acting in
real space indicate time-reversal and $\pi$-rotation around in-plane
symmetrical axis of the hexagonal lattice of localized spins,
respectively. This symmetry analysis was previously clarified in
theoretical models \cite{ChengOka}.

The graphene-like spectrum of 2D magnons is characterized by the
appearance of two valleys --- $K$ and $K'$ --- of the hexagonal
Brillouin zone. The valley index can serve as a quantum number
\cite{RycTwo,MakMcG,XuYao,PanLi,Niu}, and it has been already
demonstrated in electron systems \cite{ShaYu,VitNez}, and more
recently also in 2D photonic \cite{NohHua} and phononic \cite{LuQiu}
crystals that it can form a basis for quantum information transfer.
However, it is very difficult to use the valley degree of freedom
for 2D magnons. On one hand, the magnons near $K$ $(K')$ have large
momentum and no charge, and thus the usual magnon manipulation
methods such as FM resonance \cite{ChuVas} or electric, magnetic, or
optical methods \cite{ShaYu,VitNez,LuQiu,NohHua} can not easily
detect the two valleys, not even speaking about their difference. On
the other hand, the concept of Fermi energy is absent for bosonic
magnons, and thus one can only thermally excite the (THz) valley
magnons \cite{JinKim} simultaneously with other low-frequency modes.

Instead, in order to utilize the valley degree of freedom in 2D
magnets, we turn to the concept of topology
\cite{XiaoYao,ShiYam,SuiChen,Gorbachev,ZyuKov,ChengOka}. It is
attractive since the topological magnon transport is insensitive to
sample defects. The topological current is in principle dictated by
the crystal symmetry \cite{ShiYam,SuiChen,Gorbachev}, which is
difficult to change in 3D systems. The momentum-dependent
topological magnon current, odd under ${\cal
T}\mathcal{C}_r$ but even under ${\cal I}$
\cite{XiaoChang,ChengOka,OwePRB}, is strictly zero in pristine
magnon systems respecting both ${\cal I}$ and ${\cal
T}\mathcal{C}_r$ -- typical FM Heisenberg systems
\cite{PerBan,OwePRB}. However, if ${\cal I}$ is broken for magnons
in FM insulators, ${\cal T}\mathcal{C}_r$ requires the
Hall current to have the opposite values in two valleys, resulting
in a magnon valley Hall current, in analogy with electron systems
\cite{XiaoChang,WuZhou,LeeMak}. The ${\cal I}$ breaking can not be
induced by the external electric field, since the localized moments
do not change with the electric potential \cite{JiangShan}, however,
it can be induced by layer-dependent electrostatic doping
\cite{JiangLi} in bilayer magnets.

In this work, we explore the topological valley transport of magnons
under $\cal I$ breaking in bilayer Heisenberg FM insulators. We
propose that the pure magnon valley current can be detected via the
inverse valley Hall effect, and discuss a novel concept of valley
Seebeck effect for 2D magnets. We perform calculations for the
specific example of bilayer CrBr$_3$ \cite{ZhangShang,ChenSun},
where $\cal I$ of magnons ($\sim$1.8~THz near two valleys) can be
broken by electrostatic doping. The results show how the valley
degree of freedom can be manipulated in 2D magnets and opens the way
of using it for information transfer --- THz magnon valleytronics.

We organize the paper as follows. In Sec.~II, we introduce the
system Hamiltonian and magnon bands. In Sec.~III, we show how to
generate the topological magnon valley current. In Sec.~IV, we
demonstrate how to detect the topological magnon valley transport.
In Sec.~V, we show the material realization. In the final section,
we present the Discussion and Conclusions.

\section{II.~System Hamiltonian and Magnon Bands}

\begin{figure}
\centerline{\includegraphics[width=8cm]{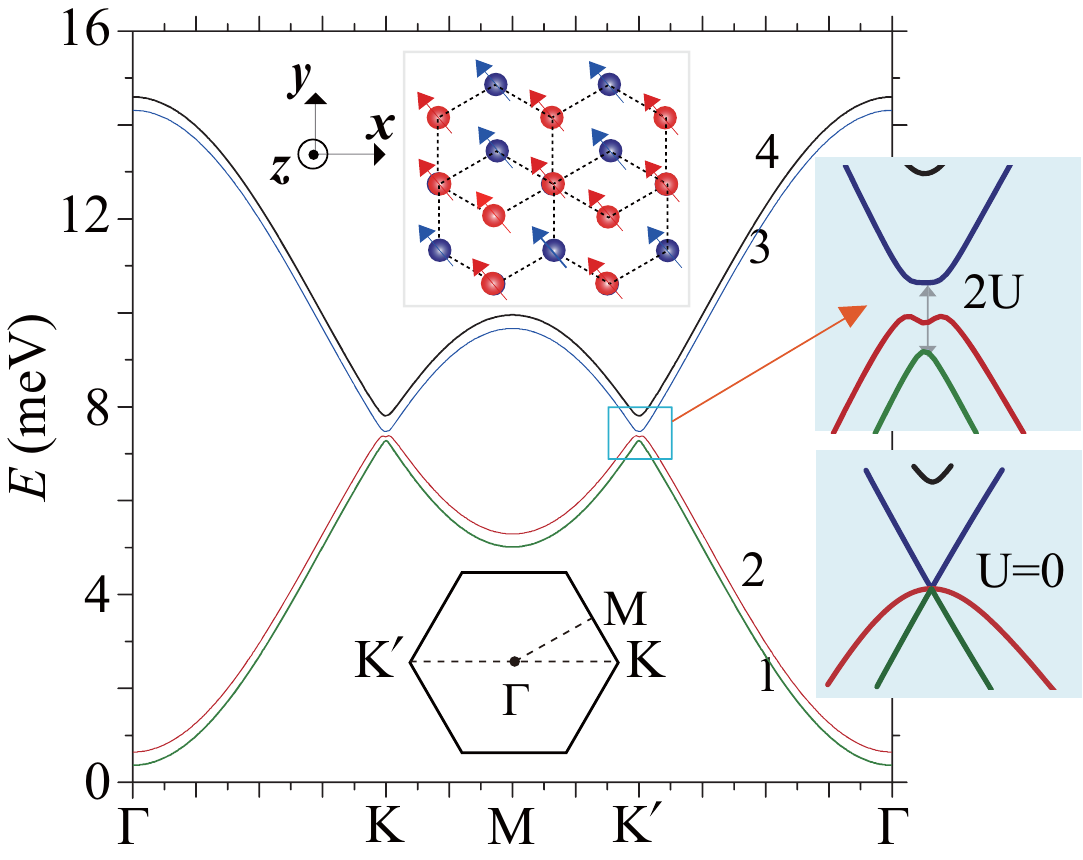}} \caption{Magnon
spectrum of bilayer CrBr$_3$ with $J=1.55$~meV, $J_z=0.2$~meV,
$S=3/2$, $\lambda=0.05J$, $a=3.72$~{\AA}, $d_z=6.11$~{\AA}
\cite{PerBan,RichWeb,ZhangShang,McGuire}, assuming $U=0.1$~meV. The
top (bottom) inset sketches the structure in real (reciprocal)
space. The right inset enlarges the gapped Dirac bands near
$7.5~{\rm meV}\simeq1.8~{\rm THz}$ (The case $U=0$ is plotted for
comparison).}
\end{figure}

We consider a bilayer collinear FM insulator on a usual AB-stacked
honeycomb lattice with the magnetic anisotropy perpendicular to the
hexagon plane, i.e. spins on $\mu$ = A$\alpha$, B$\alpha$
sublattices in the ground state satisfy ${\bm S}_{\rm A\alpha}={\bm
S}_{\rm B\alpha}=S{\bm \hat{z}}$, where $\alpha=\pm$ indicates the
top (bottom) layer, see Fig.~1. A general Heisenberg Hamiltonian
\cite{GongZhang,BurMand,GibKop,PerBan,ChenChu} with breaking of
$\cal I$ reads
\begin{equation}\label{Lattice-Ham}
\begin{split}
\hat{\cal H}=&-\sum_{\langle i,j\rangle}J_{ij}{\bm S}_i \cdot{\bm
S}_j -\lambda\sum_{\langle i,j\rangle}S_i^zS_j^z +\sum_i\alpha U{\bm
\hat{z}}\cdot {\bm S}_i.
\end{split}
\end{equation}
The first term represents the magnetic exchange interaction with
$J_{ij}=J~(J_z)>0$ for intralayer (interlayer) nearest-neighbour
magnetic moments. The second term indicates the anisotropic FM
exchange with $\lambda>0$ \cite{GongZhang}. The last term $H_U$
indicates $\cal I$ breaking [discussed in Eq.~(\ref{HU-term})]. By
applying low magnetic field, the FM ground state can be stabilized.
The essential physics will not be altered when the second- or
third-nearest neighbor exchange interactions are included.
In addition, it is intuitive to identify the symmetry
of ${\cal T}\mathcal{C}_r$ in real space. A rigorous approach to
prove ${\cal T}\mathcal{C}_r$ of the system requires expanding
Eq.~(\ref{Lattice-Ham}) to the quadratic order \cite{ChengOka}.

Neglecting the magnon-magnon interactions in
Eq.~(\ref{Lattice-Ham}), the Holstein-Primakoff (HP) transformation
\cite{PerBan} reads $S_{i\mu}^+\equiv
S_{i\mu}^x+iS_{i\mu}^y\simeq\sqrt{2S}c_{i\mu}$, $S_{i\mu}^-\equiv
S_{i\mu}^x-iS_{i\mu}^y\simeq\sqrt{2S}c_{i\mu}^\dag$,
$S_{i\mu}^z=S-c_{i\mu}^\dag c_{i\mu}$. In the A+, B+, A-, B- basis,
the Bloch Hamiltonian after HP transformation is expressed as
\begin{equation}\label{Bloch-Ham}
\begin{split}
H_{\bm k}=&S\left({\begin{array}{cccc}
    J_z+\gamma+U   &-Jf_{\bm k}  &0  &-J_z \\
    -Jf^*_{\bm k}  &\gamma+U   &0   &0 \\
    0  &0  &\gamma-U  &-Jf_{\bm k} \\
    -J_z   &0    &-Jf^*_{\bm k}  &J_z+\gamma-U \\
  \end{array}} \right),
\end{split}
\end{equation}
where $f_{\bm k}=\exp(-ik_ya/2)\left[2\cos(\sqrt3
k_xa/2)+\exp(i3k_ya/2)\right]$ ($a$ is the hexagon side length) and
$\gamma=3(J+\lambda)$.

We now introduce the Pauli matrices in the sublattice
($\bm{\sigma}$) and layer ($\bm\tau$) spaces, considering A$\alpha$
and the top layer (B$\alpha$ and the bottom layer) as an up (down)
pseudospin, respectively. The index $\xi=\pm$ denotes two valleys
$K$ and $K'$ at $(k_x, k_y)=[{4\pi}/{(3\sqrt3a)}](\pm1,0)$.
Expanding Eq.~(\ref{Bloch-Ham}) near $K$ and $K'$, we obtain the
effective Hamiltonian as
\begin{equation}\label{Ham-near valleys}
\begin{split}
H_{\bm q}=\left({\begin{array}{cc}
    H_{\bm q}^K &0  \\
    0   &H_{\bm q}^{K'}\\
  \end{array}} \right),~~
H_{\bm q}^\xi=\tau_0h_{\bm
q}^\xi+U_\gamma\tau_0\sigma_0+\frac{J_zS}{2}\Gamma,
\end{split}
\end{equation}
where $\bm q\equiv\bm k -K~(K')$, $U_\gamma\equiv(\gamma+\alpha
U)S$, $h_{\bm q}^\xi=\hbar \upsilon(\xi q_x\sigma_x+q_y\sigma_y)$
with $\upsilon=(3a/2)JS$, and
$\Gamma=\tau_0\sigma_0+\tau_z\sigma_z-\tau_x\sigma_x+\tau_y\sigma_y$
with $\tau_0$ ($\sigma_0$) describing the pseudospin identity
matrix.

We notice that there are two kinds of descriptions for
time-reversal symmetry of quantum magnets in the literature. One,
used by Ref. \cite{ChengOka}, is to define $\cal T$ that can
directly operate in lattice space of localized spins (as stated
above). Another description \cite{Owerre-JPCM,OwePRB} is to use
$\hat{\cal T}$ that operates only in Bogoliubov
Hamiltonian~(\ref{Bloch-Ham}) or (\ref{Ham-near valleys}).
For the former, one can combine ${\cal T}\mathcal{C}_r$
with the Heisenberg equation of motion to solve the problem of spin
precession \cite{ChengOka}. For the latter, an ordered state must be
assumed \cite{Owerre-JPCM} when only the properties of magnons (HP
bosons, collective excitation) are considered, and then $\hat{\cal
T}$ has a specific matrix form, similar to the one for fermion
systems \cite{SuzAndo}. For magnons, the symmetry analysis
performed by $\hat{\cal T}$ and ${\cal T}\mathcal{C}_r$ from
different perspectives \cite{ChengOka,OwePRB} lead to the same
results. For the system we study, $\hat{\cal T}$ that interchanges
two valleys is found as
\begin{equation}\label{TRS-Operator}
\hat{\cal T}=\left({\begin{array}{cc}
    0 &\tau_z\sigma_z  \\
    \tau_z\sigma_z   &0\\
  \end{array}} \right){\cal C}={\hat{\cal T}}^{-1},
\end{equation}
where $\cal C$ is the operator of complex conjugation. The relation
$\hat{\cal T}{\cal H}_{\bm q}{\hat{\cal T}}^{-1}={\cal H}_{\bm q}$
confirms the time-reversal invariance of Bogoliubov Hamiltonian,
leading to the energetically degeneracy of two valleys.
The relation ${\hat{\cal T}}^2=1$ just reflects the
spinless properties of magnons. The complex-conjugate property of
Bloch functions between two valleys further ensures the validity of
$\hat{\cal T}$ \cite{SuzAndo}.

For convenience, we set $\hbar=1$, $\upsilon=1$ below. For $U=0$,
the eigenvalues of Eq.~{(\ref{Ham-near valleys})} read $E_n^0=\gamma
S+\varepsilon_n^0$, where $n=1$ to 4 indexes the subband. Two magnon
modes in each valley are massless Dirac modes (Fig.~1),
$\varepsilon_{1,3}^0=\mp{\bm q}$, and the other two modes have a
gap, $\varepsilon_{2,4}^0=J_zS\mp[q^2+(J_zS)^2]^{1/2}$. Therefore, a
triple degeneracy \cite{OwePRB} exists at $K~(K')$ point $(q=0)$.
For $U\neq0$, the eigenvalues are $E_n=\gamma S+\varepsilon_n$,
$\cal I$ is broken, the Dirac magnon subbands are gapped (Fig.~1),
and the triplet degeneracy in each valley is lifted although the two
valleys are energetically degenerate. In contrast, there is no
counterpart of this model in 2D electron systems
\cite{XuYao,PanLi,Niu,ShaYu,VitNez,McCann}.

By using perturbation theories \cite{VanVleck,Lowdin,SuppMater}
under $U/J_z\ll1$, we can find the eigenvalues $\varepsilon_n$ as
\begin{equation}
\varepsilon_{1}\simeq\frac{\Delta_-}{2}-\sqrt{2
q^2\cos^2\frac{\varphi}{2}+\frac{\Delta_+^2}{4}},~
\varepsilon_4\simeq \varepsilon_4^0+\frac{U^2S}{2J_z}
(2-\cos\theta),
\end{equation}
where $\Delta_\pm=[J_z-(J_z^2+U^2)^{1/2}\pm U]S$,
$\cos\varphi=U/(J_z^2+U^2)^{1/2}$,
$\cos\theta=J_z/(q^2+J_z^2)^{1/2}$. Here, $\varepsilon_{2,3}(q)$ are
not presented due to their complexity (Supplemental Material (SM)
\cite{SuppMater}). Specifically, the energies at $K~(K')$ read
\begin{equation}
\varepsilon_{1,3}^0=\mp
US,~\varepsilon_{2,4}^0=\left(J_z\mp\sqrt{J_z^2+U^2}\right)S,
\end{equation}
agreeing with the numerical result in Fig.~1.

\section{III.~Generation of topological magnon valley current}

We consider a rectangular sample with width $W$ and length $L$. By
applying energy-flux quantum theory \cite{KatNag,HanLee,Onose} to
noninteracting magnons, we obtain the average energy current density
as
\begin{equation}
{\bm J}_\varepsilon=\frac{1}{2V}\sum_{\bm k}u^\dag_{m\bm
k}\left(\frac{{\partial{\cal H}_{\bm k}}^2}{\partial{\bm
k}}\right)_{mn}u_{n\bm k},
\end{equation}
where $V=WLd_z$ ($d_z$ is the thickness) is the sample volume, and
$u_{n\bm k}$ is the Bloch wave function for the $n$-th subband.
Using the Kubo formula acting on ${\bm J}_\varepsilon$
\cite{KatNag}, the thermal Hall conductivity of magnons can be
derived as $\kappa_{xy}=-(k_{\rm B}^2T/\hbar V)\sum_{n{\bm k}}{\cal
C}_{n\bm k}\Omega_{n\bm k}$, where the sum is all over the first
Brillouin zone, ${\cal C}_{n\bm k}=(1+f_{\rm B})[\ln(1+1/f_{\rm
B})]^2-(\ln f_{\rm B})^2- 2{\rm Li}_2(-f_{\rm B})$, with $f_{\rm
B}=[\exp({\varepsilon_{n\bm k}/k_{\rm B}T})-1]^{-1}$ denoting the
Bose-Einstein distribution and ${\rm Li}_2$ being the dilogarithm
function. The Berry curvature here is determined by
\cite{XiaoChang} ${\bm \Omega}_{n\bm k}\equiv{\bm \nabla}_{\bm k}
\times\langle u_{n\bm k}|i{\bm\nabla}_{\bm k}|u_{n\bm k}\rangle=
\Omega_{n\bm k}\hat{{\bm z}}$, where $\hat{{\bm z}}$ is the unit
vector in the out-of-plane direction.

\begin{figure}
\centerline{\includegraphics[width=8cm]{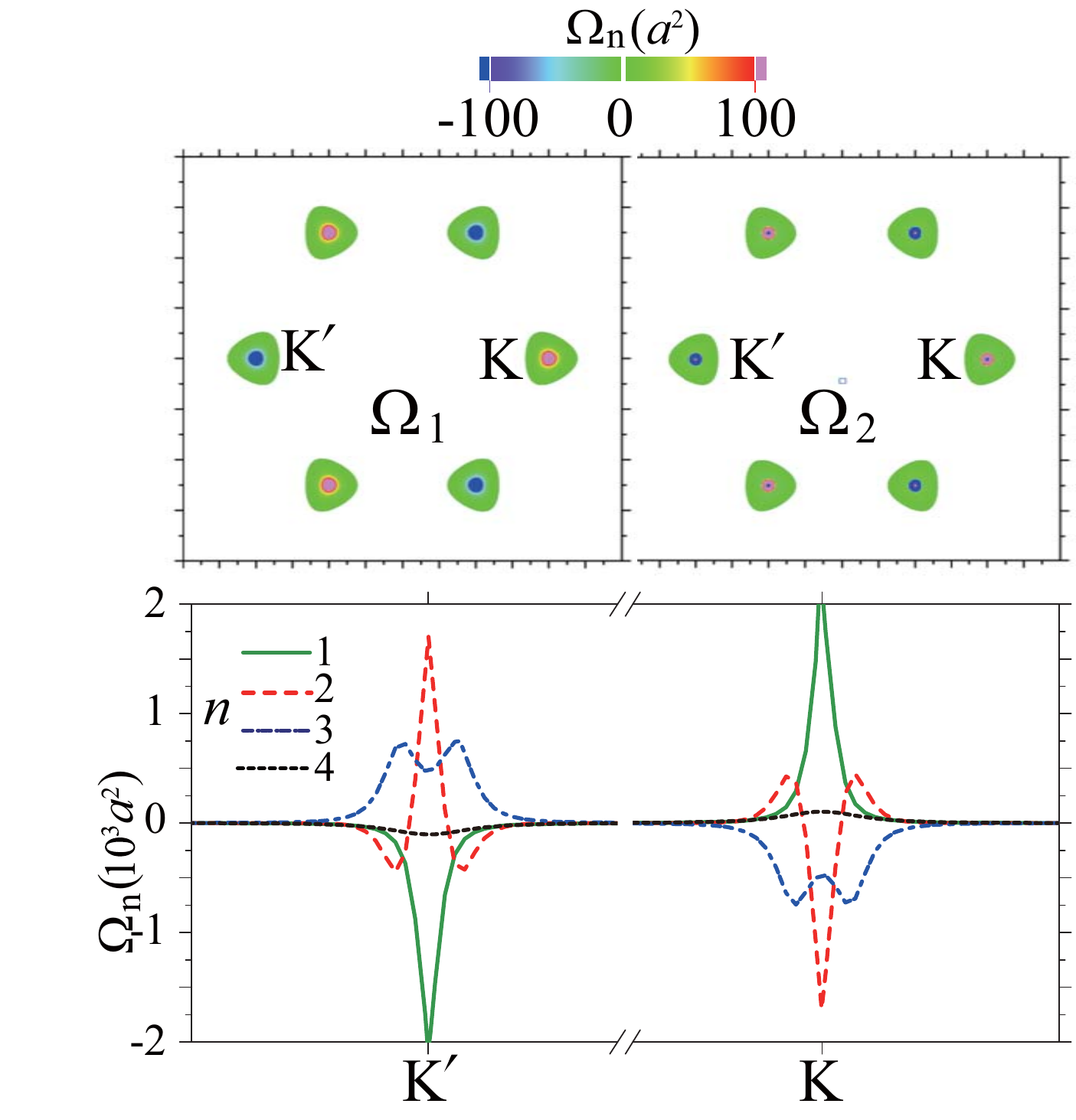}} \caption{Berry
curvature in the $(k_x,k_y)$ plane (top) and along the $K'$-$K$
direction (bottom), corresponding to the gapped Dirac magnon band in
Fig.~1.}
\end{figure}

Because the magnonic system has $\hat{\cal T}$, the magnons near $K$
and $K'$ can feel the opposite orbit pseudomagnetic field reflected
in $\Omega_{n\bm q}^K=-\Omega_{n,-\bm q}^{K'}$, as shown in Fig.~3.
Our detailed calculations \cite{SuppMater} reveal the local
conservation law of the Berry curvature \cite{XiaoChang}
$\sum_{n=1}^4\Omega_{n\bm k}=0$. We have $\kappa_{xy}=0$ because the
contributions from $K$ and $K'$ cancel each other, however, a pure
valley current of magnons is generated. To distinguish the
difference between $K$ and $K'$, it is necessary to define a magnon
valley Hall conductivity as $\kappa_{xy}^{\rm
v}\equiv\kappa_{xy}^K-\kappa_{xy}^{K'}$. It has the form
\begin{equation}
\begin{split}
\kappa_{xy}^{\rm v}=-\frac{k_{\rm B}^2T}{\hbar V}\sum_{n{\bm
q}}~{\cal C}_{n\bm q}~(\Omega_{n\bm q}^K-\Omega_{n\bm q}^{K'}),
\end{split}
\end{equation}
where the summation over $\bm q$ around valley $K$ ($K'$) runs over
the half of the first Brillouin zone. As we concern in
Eq.~(\ref{Ham-near valleys}), the pure valley transport occurs due
to $\Omega_{n\bm q}^K = -\Omega_{n\bm q}^{K'}$.

\section{IV.~Detection of topological magnon valley transport}

The signal of topological valley transport $\kappa_{xy}^{\rm v}$ can
be detected via the inverse magnon valley Hall effect, as sketched
in Fig.~3(a). For thermal magnon transport, the directly observable
qualities are thermal flux density $j_\varepsilon$ and temperature
difference \cite{KatNag}, {\it e.g.} $\Delta T_{\rm L}$ and $\Delta
T_{\rm R}$ in Fig.~3(a). From the well-known Onsager relation
\cite{BauerSaitoh}, we define a nonlocal thermal resistivity as
$\rho_{\rm NL}\equiv\Delta T_{\rm R}/j_\varepsilon$. By using the
appropriate boundary conditions (SM \cite{SuppMater}), we
self-consistently derive $\rho_{\rm NL}$ as
\begin{equation}\label{Nonlocal-R}
\begin{split}
\rho_{\rm NL}\approx\frac{W}{2\ell_{\rm v}}\frac{(\kappa_{xy}^{\rm
v})^2}{(\kappa_{xx})^3}{\exp\left(-\frac{L}{\ell_{\rm v}}\right)},
\end{split}
\end{equation}
where $\kappa_{xx}=j_\varepsilon/\Delta T_{\rm L}$ is the local
thermal conductivity. Note that Eq.~(\ref{Nonlocal-R}) recovers the
formula known for spin or valley Hall systems for electrons
\cite{ShiYam,AbanShy}.

We emphasize that the setup of Fig.~3(a) is
specifically designed to measure the contribution of the valley Hall
effect. Indeed, no signal can be detected along $x$ direction for
pure magnon valley Hall current we focus. This means that one can
add a detection along $x$ direction to exclude the normal thermal
Hall effect of magnons. To exclude the contribution from phonons and
electrons to the thermal Hall effect, we present the following
argument. First, we can ignore the electron transport in the
insulating regime. Beyond this regime, the electron contribution to
the thermal Hall current can be excluded by adjusting the Fermi
energy, because the band topology strongly depends on the Fermi
energy \cite{ShiYam,SuiChen}. Second, the contribution from phonons
to the thermal Hall current is absent in ordinary materials
(including 2D magnets concerned here) because the observable phonon
Hall effect usually needs special interaction mechanisms
\cite{LiFauq}. Third, the contribution of electrons and phonons to
$\kappa_{xx}$ (not Hall current) may be comparable to that of
magnons at low temperatures \cite{SuppMater,Onose}, but only brings
about a certain reduction of $\rho_{\rm NL}$ (This has been taken
into account in the estimates of Sec.~V).Thus, the contribution of
the pure magnon valley Hall effect can be quantified by using the
design of Fig.~3(a).

\begin{figure}
\centerline{\includegraphics[width=8cm]{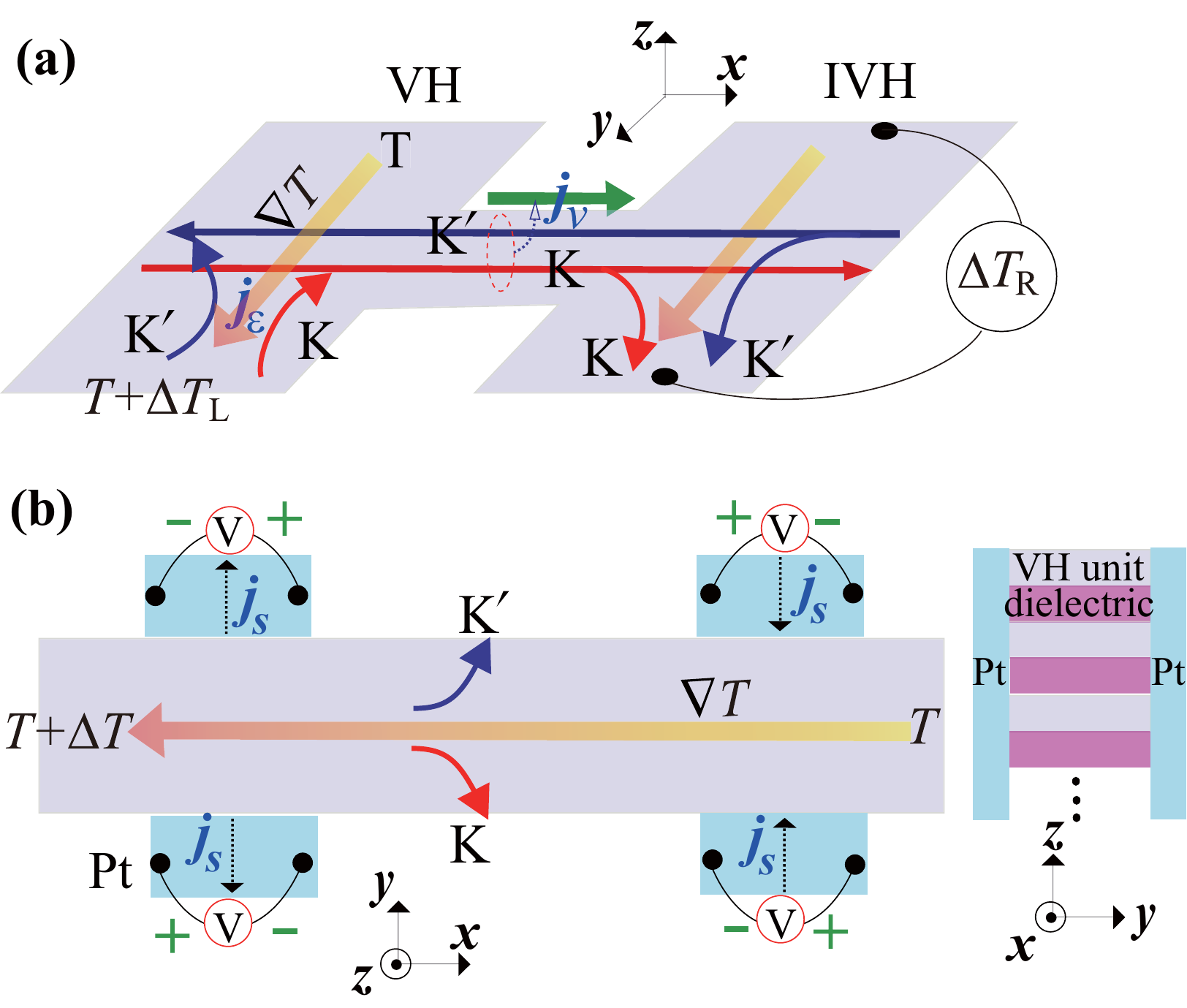}} \caption{(a) A
Hall bar to thermally detect a net pure valley Hall (VH) current
${\bm j}_{\rm v}$ via the inverse VH (IVH) effect,
analogous to the detector for the electron case
\cite{ShiYam,VitNez}. The
valley Hall current along $x$ direction comes from the left bar when
magnon energy current ${\bm j}_\varepsilon$ flows along $y$
direction and is detected on the right bar. (b) Left: A device of
valley Seebeck effect for magnons. Inverse spin Hall (ISH) effect
\cite{UchidaXiao,BauerSaitoh} occurs in Pt metallic contact
(optionally), where ${\bm j}_s$ is the spin current.
Right: A proposed valley-Seebeck integrated system
composed of VH units separated by nonmagnetic dielectric layer ({\it
e.g.} layered boron nitride \cite{JiangLi}). This setup is suggested
to increase the measured signal.}
\end{figure}

An alternative method to detect $\kappa_{xy}^{\rm v}$ is by using
heat-to-charge conversion \cite{BauerSaitoh}. We propose a concept
of valley Seebeck effect, as sketched in Fig.~3(b), where Pt
contacts (optionally) are added to the sample, and magnons from $K$,
$K'$ accumulate at the opposite sides. Consequently, spin current
${\bm j}_s$ can be induced in Pt due to s-d interaction (SM
\cite{SuppMater}) at the FM insulator/Pt metal interface
\cite{BauerSaitoh,UchidaXiao,CornPeters}. The inverse spin Hall
(ISH) electric field in Pt is determined by
\begin{equation}\label{ISH}
{\bm E}_{\rm ISH}=(\rho\theta_{\rm SH}){\bm j}_s\times{\bm s},
\end{equation}
where $\rho$ and $\theta_{\rm SH}$ indicate the electric resistance
and spin Hall angle of Pt, ${\bm s}$ is an out-of-plane spin
polarization vector. The ISH voltage is in principle determined by
\cite{SuppMater} $V_{\rm ISH}\propto j_s\propto\Delta
T\sinh(x/\lambda_m)$, where $x=0$ is at the center of the CrBr$_3$
sample and $\lambda_m$ is the magnon relaxation length. Because the
temperature difference between electrons in Pt and magnons in FM
insulator changes sign if the Pt contact is moved from left to right
in Fig.~3(b), $j_s$ and the ISH voltage change signs accordingly
\cite{UchidaXiao,SuppMater}. In the absence of valley
Seebeck effect in Fig.~3(b), we judge that the contributions from
normal spin Seebeck effect to ISH signal along $+y$ and $-y$
directions should in principle cancel out each other (independent of
$x$) due to the transport symmetry. In this case, the net magnon
current is in $x$ direction (along the temperature gradient), and no
net magnon current in $y$ direction is converted into spin current
in Pt leads under $\partial T/\partial y=0$ in the magnet. Unlike
the usual 3D case \cite{UchidaXiao}, the transport in $z$ direction
for the ultrathin 2D case here is negligible.

We now discuss deeply the detection in Fig.~3(b) as
follows. First, an ideal way to make this setup requires growing
metal leads attaching at the boundary of the 2D system. By this
means, one can truly detect the 2D in-plane transport (excluding the
out-of-plane contribution). By contrast, the recent experimental
setups \cite{LiuPei,XingQiu} were mainly based on the up-down
configurations of metal leads and thicker 2D magnets (sub-10~nm
thick), but physically the out-of-plane spin pumping probably
dominates. Second, the ISH voltage is proportional to $\theta_{\rm
SH}$ according to Eq.~(10). Due to the rather-small thickness of the
2D magnet, $\theta_{\rm SH}$ might be very small if the usually
thicker metal leads are used due to the size effect. This would be
an obstacle to the detection of the 2D in-plane magnon transport,
although making metal leads thin enough is not impossible
(technically, it is very difficult). To overcome this problem, we
suggest that $\theta_{\rm SH}$ and valley Seebeck signal can be
greatly enhanced by vertically integrating valley Hall units, as
sketched in Fig.~3(b) (right panel). In this case, the size of
contact surfaces between metal leads and 2D magnets is significantly
increased. By assuming that the metal leads have the similar-sized
thickness with the integrated valley-Hall system, we still use the
typical value of $\theta_{\rm SH}$ to give estimations below.

In addition, it is assumed that the length and the
width of the 2D magnet are on the order of $1~\mu$m (see detailed
estimation in Sec.~V), and the temperature differences $\Delta
T_{\rm L}$ in Fig.~3(a) and $\Delta T$ in Fig.~3(b) should be much
lower than the basic temperature $T$ and are on the orders of
$10^{-3}\sim1$~K, which is available with the application of current
experimental techniques \cite{BauerSaitoh,LiuPei,XingQiu}. The
measured signal can be effectively enhanced by increasing the
temperature difference.

\section{V.~Material Realization}

\begin{figure}
\centerline{\includegraphics[width=8cm]{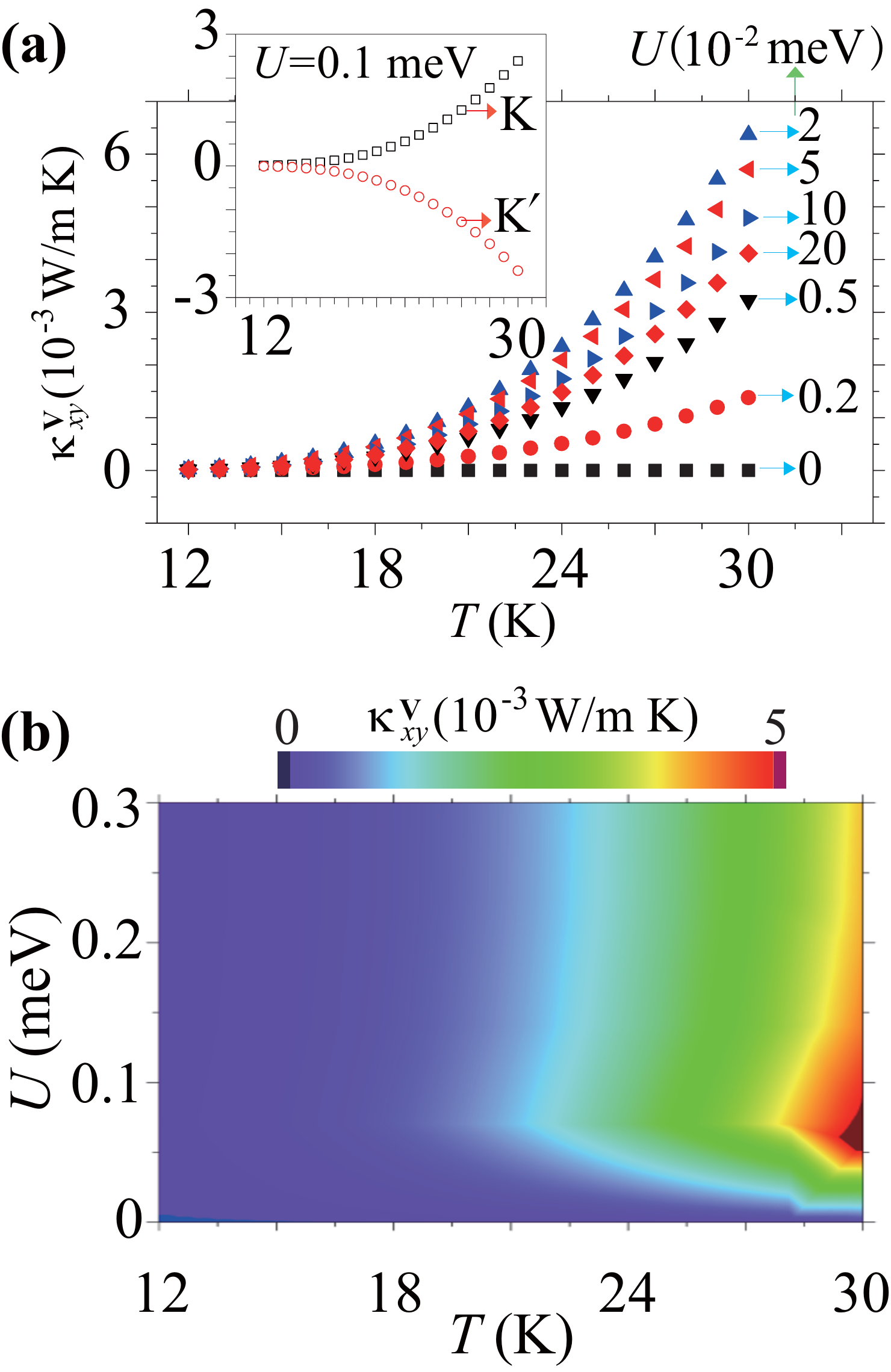}} \caption{ Valley
Hall conductivity (a) as a function of temperature under the
influence of $U$ and (b) in the $(U, T)$ plane. The other parameters
are taken from CrBr$_3$ (see the caption in Fig.~1).}
\end{figure}

\begin{figure}
\centerline{\includegraphics[width=8cm]{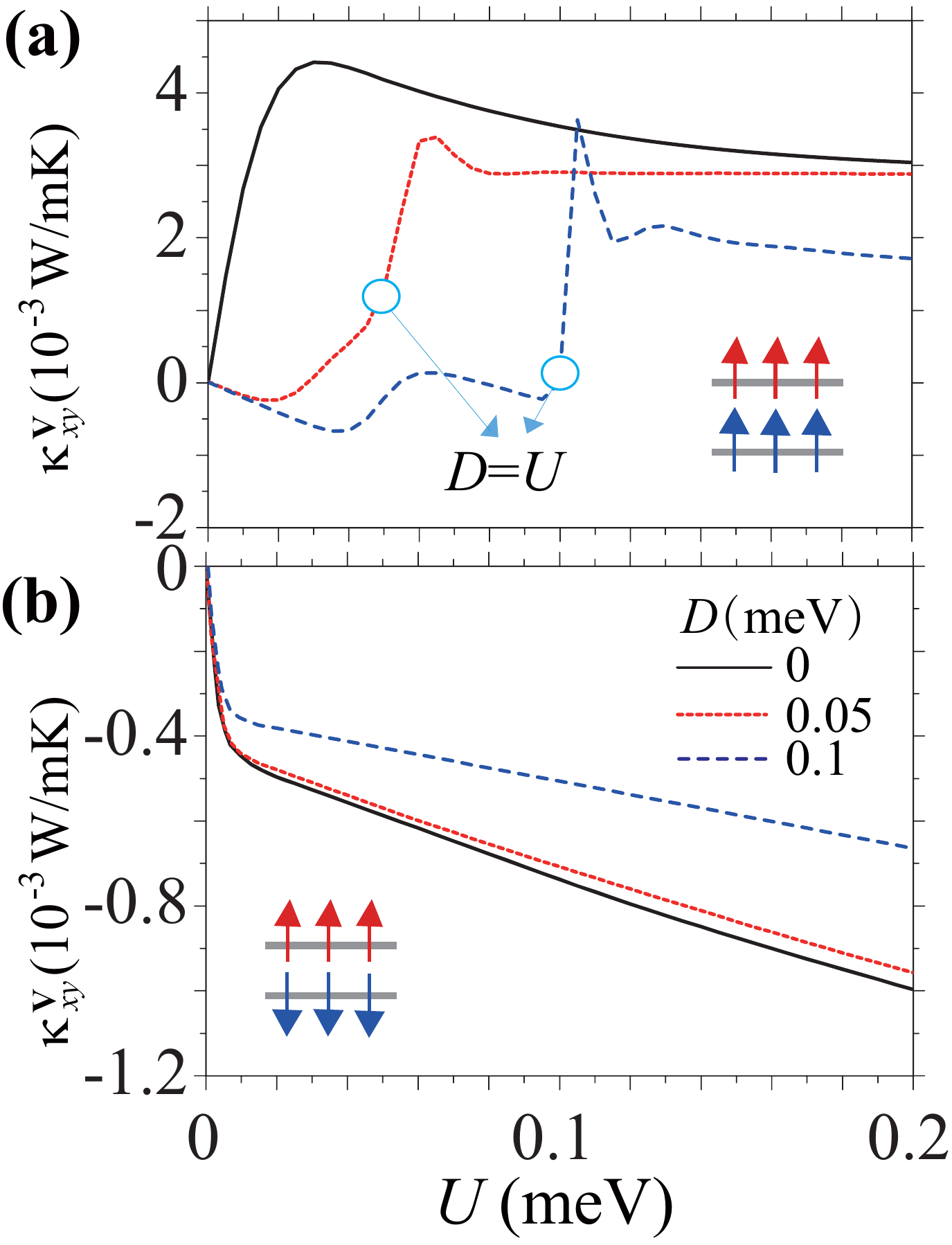}} \caption{Valley
Hall conductivity as a function of potential $U$ under the influence
of DMI in (a) FM and (b) LAF samples at $T=28$~K. The solid, dot,
dashed lines represent the results of $D=0$, 0.05, 0.1 in units of
meV, respectively. The other parameters from CrBr$_3$ in Fig.~1 are
taken. Pure valley current $\kappa_{xy}^{\rm K}=-\kappa_{xy}^{\rm
K'}$ \cite{SuppMater} is only present for $D=0$ in both (a) and
(b).}
\end{figure}

Our proposal for topological valley Hall effect of magnons can be
experimentally detected in a number of 2D FM insulators. One
possibility is bilayer CrBr$_3$, which has an electronic band gap
about $1.4$~eV \cite{WebYan,WangEye} and $S=3/2$ for each Cr$^{3+}$
ion \cite{PerBan}. According to experimental results
\cite{PerBan,WebYan,ZhangShang}, the Hamiltonian~(\ref{Lattice-Ham})
without $H_U$ indeed captures the intrinsic magnon physic of this
material, and quantum fluctuations are not as important as that in
spin $1/2$ systems \cite{ChengOka}. In addition, corrections from
the single-ion magnetic anisotropy term $H_{\cal A}={\cal
A}\sum_{\langle i,j\rangle}(S_i^z)^2$ shift up the magnon bands by a
small energy $2{\cal A}$ below $0.1$~meV \cite{AbrJas}. This effect
is not relevant for topological transport dominated by high-energy
THz magnons near two valleys.

By applying the electrostatic doping technique \cite{AhnCH} to bilayer
CrBr$_3$, the localized moments in each layer can be continually
tuned (more than $20\%$ electron or hole doping was achieved in
CrI$_3$ \cite{JiangLi}). For a typical doping case, the two
monolayers are doped equally with opposite charges, and the localized
moments in the FM ground state can be described as ${\bm
S}_{i,\alpha}=(S-\alpha\Delta S){\bm \hat{z}}$. One advantage of
this doping is that the variation of parameters $J_\bot$, $\lambda$
in Eq.~(\ref{Lattice-Ham}) can be safely ignored \cite{JiangLi}. The
$\cal I$ breaking term $H_U$ has the specific form
\begin{equation}\label{HU-term}
H_U=\sum_i\alpha U S_i^z~~~{\rm with}~~~U=\Delta S \ .
\end{equation}
The doping also creates interlayer electric field. However, we
expect that it has a minor contribution to $\Delta S$
\cite{JiangShan} because the two insulating layers are weakly
coupled.

Figure~4 shows the results of the calculation of $\kappa_{xy}^{\rm
v}$ for bilayer CrBr$_3$ for different $U$ and $T$. They indicate
that $\kappa_{xy}^{\rm v}$ is lower than 0.1~mW/m K below 12~K
because the Berry curvature for the dominant low-energy magnons away
from two valleys is nearly zero (Fig.~2). Near $T=30$~K (Curie
temperature $T_{\rm C}=34$~K \cite{ZhangShang}), $\kappa_{xy}^{\rm
v}$ can exceed 6~mW/m K, which is about one order of magnitude
greater than the Hall signal reported in 3D FM insulators
\cite{Onose} demonstrating the advantage of Dirac magnons here. We
assume a typical value of
$\kappa_{xy}/\kappa_{xx}\sim5\times10^{-3}$ (see SM
\cite{SuppMater}, where $\kappa_{xx}$ contributed from electrons,
phonons and magnons are discussed), which agrees with the latest
experimental detection in other 2D magnets \cite{XingQiu,WangXS}. By
fixing $L=4W=2\ell_{\rm v}$ ($\ell_{\rm v}\sim1.0~\mu\rm m$ is
expected from valleytronic experiments
\cite{ShiYam,SuiChen,Gorbachev}), $\rho_{\rm NL}$ is estimated from
Eq.~(\ref{Nonlocal-R}) to be about $4\times10^{-6}$~m K/W which in
principle is experimentally detectable \cite{UchidaXiao}. The ISH
voltage in Fig.~4 is estimated from Eq.~(\ref{ISH}) as $V_{\rm
ISH}\sim0.06~\mu{\rm V/K}$ (SM \cite{SuppMater}) by using the
typical parameters \cite{UchidaXiao} $\theta_{\rm SH}=0.0037$,
$\rho=0.91\mu\Omega{\rm m}$ in Pt. Above $T_{\rm C}$, the detection
signal should decrease drastically due to the enhanced magnetic
disorder \cite{Onose}.

Permitted by symmetry in hexagonal structure, there may exist the
Dzyaloshinskii-Moriya interaction (DMI)
\cite{KimOchoa,ChenChu,OwePRB}, $H_{\rm
DM}=[D/(3\sqrt3)]\sum_{\langle\langle
ij\rangle\rangle}\nu_{ij}\hat{{\bm z}}\cdot({\bm S_i}\times{\bm
S}_j)$ with $\nu_{ij}=+1$~(-1) if the exchange between two
next-nearest spins is clockwise (anticlockwise). DMI can induce
magnon Hall effect \cite{OwePRB} by breaking $\cal T$ but does not
generate topological difference between two valleys. To clarify the
influence of DMI, we calculate $\kappa_{xy}^{\rm v}$ as a function
of $U$ in Fig.~5(a) for $D=0,~0.05,~0.1$~meV. As expected,
$\kappa_{xy}^{\rm v}$ becomes weaker under the effect of DMI, and
can be enhanced by increasing $U$. Near the circled points $D=U$,
the gap closing and reopening near one valley occur (SM
\cite{SuppMater}), and $\kappa_{xy}$ changes drastically due to the
sign change of Berry curvature.

Moreover, we consider the influence of layered antiferromagnetic
(LAF) order \cite{ChenSun}, which means each layer is FM while the
FM orientation is opposite between two layers. For the LAF system,
$H_U$ in Eq.~(\ref{Lattice-Ham}) should be replaced by $H_U=U\sum_i
S_i^z$ (in the ground state, $S_i^z$ is opposite for two layers).
The results shown in Fig.~5(b) indicate that the layer polarity term
$H_U$ is still useful to strengthen the signal of $\kappa_{xy}^{\rm
v}$, reaching 1~mW/m K at $T=28$~K for a weaker DMI. Compared with
the FM bilayer case, the almost opposite Berry curvature
\cite{SuppMater,OwePRB} of magnon bands for the LAF case hinders the
enhancement of topological valley current. To induce $H_U$, the
perpendicular magnetic field \cite{JiangShan} is feasible besides
doping. As $U$ increases, a transition of LAF to FM \cite{JiangShan}
should happen, and a stronger topological valley signal is
detectable.

\section{VI.~Discussion and Conclusions}

The picture of topological valley transport of magnons discussed in
bilayer CrBr$_3$ also applies to multilayer FM CrBr$_3$
\cite{RichWeb,ZhangShang,McGuire}, bilayer or multilayer
CrBr$_x$I$_{3-x}$ \cite{AbrJas} or CrI$_3$ \cite{JiangLi}. A
particular issue is the Gilbert damping \cite{ChuVas}, for which the
damping constant in Landau-Lifshitz-Gilbert equation is estimated as
\cite{WangXS,XingQiu,PerBag,KimOchoa} $10^{-4}$ under the effects of
the weak Rashba-type DMI (an interaction between
nearest neighbors of localized spins \cite{KimOchoa}) or disorder.
The Gilbert damping usually affects the magnon-related
transport properties by determining the magnon lifetime
\cite{RuckBrat,BauerSaitoh}, and might be weakened by improving the
sample quality and optimizing the sample scales. This damping and
our proposed effects are expected to change little in weak doping
case $U\ll J$ we focus (the system can preserve insulating
properties by adjusting gates). However, if the doped
magnet becomes metallic, the localized spin model alone probably
does not work. Additional significant effects to influence the
magnon valley Hall effect should be considered due to the
interaction between magnons and electrons/holes. It is still an open
question that whether more complicated mechanisms work for larger
$U$ or doping-induced metallic case. Certainly, it is promising in
view of application to apply the physics discussed
here to other 2D FM insulators with higher Curie temperature.

We now discuss DMI interactions. On one hand, the DMI
parameter $D$ (Fig.~5) in the two layers of the bilayer system are
considered to be of the same sign, because this DMI interaction
between next-nearest neighbors of localized spins originates from
the atomic spin-orbit couplings \cite{McGuire,AbrJas,ChenChu} and is
determined by the atomic structure of the crystal. Further
calculations indicate that, when $D$ in the two layers take the
opposite sign, the signal of $\kappa_{xy}^{\rm v}$ can not be
stronger than that in the case of $D=0$. On the other hand, the
Rashba-type DMI may be enhanced at some interfaces between the
magnet and its coupled proximity materials, by van der Waals
engineering \cite{GongZhang} or by increasing an out-of-plane
electric field \cite{KimOchoa}. If the Rashba DMI is strong enough
(approximately exceeding $J_z$), the magnon valley Hall transport is
significantly modulated due to the topological phase transitions of
energy bands, as demonstrated in similar electronic systems based on
bilayer graphene \cite{RenQiao,ZhaiJin}.

Moreover, it is still necessary to clarify the
following two points. First, for the 2D bulk system we consider,
there is no anisotropy because the magnon Hall effect is dominated
by the bulk states that are basically isotropic (as shown by the
magnon bands near two valleys in Fig.~1). Furthermore, the result
$\kappa_{yx}^{\rm v}=\kappa_{xy}^{\rm v}$ (by switching $x$ and $y$
in calculations) proves this point. However, for narrow nanorribons
with zigzag and armchair edges, the anisotropy of magnon valley Hall
conductivity can not be ignored due to the quantum confinement
effect, as also happens in the electronic case \cite{ZhaiBlanter}.
Second, inelastic many-body interactions between magnons should be
considered when the basic temperature $T$ of the system approches
$T_{\rm C}$. According to the results of Ref.~\cite{PerBan}, the
many-body interactions in crystal would renormalize the magnon bands
(the strength of the renormalization depends on temperature), but
however would not alter the symmetries of the system. In this sense,
the symmetry-protected phenomenon of topological valley transport of
magnons is robust.

In summary, we have shown that THz magnons are a new attractive
platform for valleytronics beyond fermions and form the basis for
valley-controlled magnonic applications. Our results will motivate
the experimental exploration of valley-related magnon physics in 2D
van der Waals magnets. Richer valley properties of topology and
transport can be expected under the combined effect of magnetic
order, symmetry breaking, and magnetic interactions.

\section*{Acknowledgement}

We thank T. Yu and G. E. W. Bauer for helpful discussions. The work
was supported by the National Natural Science Foundation of China
with Grant No.~61874057, the QingLan Project of Jiangsu Province
(2019), the 1311 Talent Program ``DingXin Scholar" of NJUPT (2018)
and the Jiangsu Government Scholarship for Overseas Studies.

\input{Bilayer_Magnet-Magnon_Valleytronics.bbl}

\end{document}